\documentclass[acmsmall,screen]{acmart}
\usepackage{graphicx} 

\usepackage{float}
\usepackage{amsmath}
\usepackage{xcolor}
\usepackage{subcaption}

\usepackage{dirtree}
  \usepackage{xcolor}
  \definecolor{cvrgone}{HTML}{A8452B}
  \definecolor{cvrcopied}{HTML}{2F7A55}

\usepackage{hyperref}

\usepackage{booktabs}
\usepackage{tabularx}
\usepackage{tikz}
\usepackage{array}

\newcommand{\cmark}{%
  \tikz[baseline=-0.5ex,x=1ex,y=1ex]
    \draw[line width=0.08ex,line cap=round,line join=round]
      (0,0.35) -- (0.3,0) -- (0.9,0.8);%
}

\newcommand{\emptybubble}{%
  \tikz[baseline=-0.6ex]\draw (0,0) circle (0.11);
}

\newcommand{\filledbubble}{%
  \tikz[baseline=-0.6ex]\fill (0,0) circle (0.11);
}

\usepackage{fontspec}
\usepackage{fvextra}
\usepackage{newfloat}
\DeclareFloatingEnvironment{listing}
\DefineVerbatimEnvironment{MintedVerbatim}{Verbatim}{fontsize=\small,breaklines}
\input{highlighting/style.tex}

\newenvironment{statute}[1]
  {\begin{quote}\small\textbf{#1.}\ }
  {\end{quote}}

\newcommand{\ballotmark}[1]{\ifnum#1=1 \filledbubble\else\emptybubble\fi}
\newcommand{\ballotline}[6]{#1 & \ballotmark{#2} & \ballotmark{#3} & \ballotmark{#4} & \ballotmark{#5} & \ballotmark{#6} \\}
\newcommand{\ballothead}{%
  \toprule
  \multicolumn{6}{c}{\textbf{United States Representative}} \\
  \midrule
  & \scriptsize 1st & \scriptsize 2nd & \scriptsize 3rd & \scriptsize 4th & \scriptsize 5th \\
  \midrule}

\newcommand{\statutefrac}{0.30}
\newcommand{\leanfrac}{0.67}

\title{Counting by the Book}
\subtitle{Verifying Alaska's Ranked-Choice Elections}

\author{Lucy Horowitz}
\affiliation{%
  \institution{University of California, Berkeley}
  \city{Berkeley}
  \state{California}
  \country{USA}
}
\email{lucyhorowitz@berkeley.edu}

\begin{abstract}
We present Alaska's ranked-choice elections as a case study in Formally Verified Democracy, a program to establish election results through formally checked proofs. We formalize an interpretation of Alaska's tabulation rules in Lean, verify its implementation, and prove election-specific results. Across 115 ranked-choice contests in Alaska since 2022, we identify 75 ballot records across 10 contests that were counted differently from this interpretation by a program reproducing the published results, without changing any winners. We trace these differences to overvotes and duplicate rankings and examine their legal justification, showing how formal verification can expose consequential choices in translating election law into software.
\end{abstract}



\acmDOI{}

\begin{document}

\maketitle

\section{Introduction}
An election result should come with evidence that the reported winner follows from the ballots and the rules governing their tabulation. The goal of Formally Verified Democracy is to provide such evidence in the form of a machine-checked proof. We express election rules as precise specifications and establish that the reported outcomes are permitted by those specifications. Such proofs can increase confidence in election results by making the reasoning from ballots to winners independently checkable. This paper presents Alaska's ranked-choice elections as a case study in Formally Verified Democracy. We develop a verified tabulator and election-specific proofs, but also confront a prerequisite for their use: determining which rules the tabulator should implement when the governing legal text requires interpretation.

Jurisdictions across the United States are adopting ranked-choice voting (RCV) \cite{ncsl2026rcv}, where instead of casting one vote for one candidate, voters rank multiple candidates in order of preference. In most of these elections, the winners are determined by a counting method called instant-runoff voting (IRV). The precise rules for interpreting malformed ballots, including those that contain skipped rankings, repeated candidates, or multiple candidates at the same ranking, vary across jurisdictions. Seemingly minor choices can affect how ballots are counted. This paper examines the relationship between Alaska's statutory tabulation rules and the behavior recoverable from its published election records.

Alaska adopted ranked-choice voting through Ballot Measure 2 in 2020, which added the tabulation procedure in AS 15.15.350 \cite{2020_alaska_voterinfopacket,GeneralProcedureBallot}. We abbreviate references to its subsections, for example (d)(2).

The statute was first applied in the August 2022 special general election for U.S. Representative and subsequently in the 2022 and 2024 general elections.\footnote{RCV will be used in the 2026 general election in Alaska. Also on the ballot is an initiative to repeal ranked-choice voting altogether. A similar initiative was narrowly defeated in 2024, 50.1\% to 49.9\% \cite{fisher_alaska_2026}.} The Division of Elections publishes cast vote records (CVRs) and tabulation reports that permit comparison of aggregate ballot treatments in individual rounds of each contest \cite{ElectionResultsDivision2022}. 

Voters sometimes make errors when marking their ballots. Within RCV, there are many established ways to handle such ballots \cite{pettigrewOvervotesOverranksSkips2026, helgaas2025patchwork}. As a result, the same ballot may be counted differently under one set of rules than under another. For example, consider the following ballot, reconstructed from Alaska’s 2022 general-election cast vote record:

\begin{figure}[H]
  \centering
  \small
  \setlength{\tabcolsep}{3pt}
  \renewcommand{\arraystretch}{1.0}
  \begin{tabular}{@{}l*{5}{c}@{}}
    \ballothead
    \ballotline{\textbf{Begich, Nick}}{1}{1}{1}{0}{0}
    \ballotline{\textbf{Bye, Chris}}{0}{0}{0}{0}{0}
    \ballotline{\textbf{Palin, Sarah}}{0}{0}{0}{1}{0}
    \ballotline{\textbf{Peltola, Mary S.}}{0}{0}{0}{0}{0}
    \ballotline{Write-in:}{0}{0}{0}{0}{0}
    \bottomrule
  \end{tabular}

  \caption{A ballot reconstructed from Alaska's 2022 general-election cast vote record (ballot 82311)
    \cite{alaskadivisionofelectionsCastVoteRecord2022}.}
  \label{fig:intro-ballot}
\end{figure}

This voter ranked Begich first, second, and third, followed by Palin fourth. Alaska’s regulations instruct election officials to count a repeated candidate at their highest ranking and “disregard the remaining rankings for that candidate.” Does disregarding the repeated Begich rankings make the second and third rankings blank? The distinction matters because Alaska's rules also say that two consecutive skipped rankings make a ballot inactive \cite{DuplicateRankings,GeneralProcedureBallot}. After Begich’s defeat, our statutory model (Section \ref{sec:statutory}) counts this ballot for Palin, while the model that reproduces the published results treats it as inactive and doesn't count it for any candidate.\footnote{The necessity of the latter treatment  in reproducing Alaska's results was discussed by Manuel Alvarez-Rivera in \cite{alvarez-riveraRankedChoiceVoting2023}.} 

Which set of rules were applied in Alaska's ranked-choice elections, and does this set agree with what is prescribed by the statute? The software used by the Alaska Division of Elections to tabulate ranked-choice ballots was developed by Dominion Voting Systems (now Liberty Vote) and is proprietary. Consequently, its implementation cannot be publicly formally verified against the statute or even inspected. However, the published data permit a form of black-box analysis. While we cannot prove anything about the actual software that they used, independently writing code that tabulates the votes from the ballots in the CVR and matches the published round-by-round totals allows us to infer certain properties of the implementation's behavior.

The properties of tabulation software can materially affect the outcome of elections. In Oakland, CA's November 2022 District 4 school board election, an error in how the tabulation software was configured to handle a certain class of mismarked ballots caused Nick Resnick to be declared the winner instead of Mike Hutchinson \cite{mccuneRankedChoiceBedlam2023}. An independently checked proof of the result against a specification capturing the required ballot treatment would have exposed this discrepancy before certification.

We formalize a \textit{statutory model}: an interpretation of Alaska's rules in Lean 4~\cite{lean4}, verify its implementation, and prove theorems encoding the results of specific elections (Section~\ref{sec:statutory}). Across 115 contests (some of which had only one round of tabulation), our interpretation counts 75 ballot records differently from a model that exactly reproduces all 22 published multi-round reports (the \textit{reproduction model}), while electing the same winners (Section~\ref{sec:empirical}). In Section~\ref{sec:legal} we identify the rules responsible for these ballot-level discrepancies and assess their legal justification. In Section~\ref{sec:related} we situate this case study within the landscape of formal voting theory and law.

\section{The Statutory Model}\label{sec:statutory}

This section describes the formalization of our statutory model. The formalization has three layers: statutory vocabulary, a specification of permitted outcomes, and an implementation that computes those outcomes. The \emph{statutory vocabulary} formalizes the objects of the tabulation (candidates, ballots, elections, and rounds) and some of the other formalizable vocabulary in the statute. The \emph{specification} uses this vocabulary to state which outcomes are permitted by the statute. It follows the definitions clause by clause and (d) step by step. It states permitted outcomes of a single round of tabulation (which candidate it may defeat, which it may elect, if any, and which ballots remain tabulated) and permits a sequence of defeats and a winner exactly when every step in that sequence is permitted. The \emph{implementation} is a function that computes all of the permitted outcomes of an election as a tree of defeat sequences.

The layers are held to different standards. The vocabulary and specification follow the text as closely as Lean allows because their relationship to the statute cannot be proved; it can only be judged by a reader. The implementation, on the other hand, is free to depart from the text because we prove that it computes exactly those outcomes permitted by the specification (Theorem~\ref{thm:tabulate_correct}). We nonetheless write our implementation to follow the procedure of (d) round by round so that it is easy to compare with the statute. A side-by-side comparison of the statute and the formalization is available online~\cite{horowitzFVDemocracy2026}.

Beyond the basic data model, we go into detail only on the formalizations that required significant interpretation, are ambiguous, or differ from the reproduction model. The full collection of definitions can be found in the GitHub repository\footnote{\url{https://github.com/lucyhorowitz/FVDemocracy}} at \path{FVDemocracy/Alaska/Statute/StatuteDefs.lean}.

\subsection{Statutory Vocabulary}

\subsubsection{Ballots and elections} See Listing~\ref{lst:datamodel} in Appendix~\ref{app:listings} for the Lean representations of candidates, ballots, elections, and rounds.  We represent candidates with an inductive type \verb|Cand|. A candidate may be a \verb|writeIn| or a \verb|regular| candidate, and each candidate is labeled by a unique natural number. The specific identifiers for candidates are provided by the CVR,\footnote{By AS 15.15.365(a), Alaska pools write-ins into a single candidate unless they ``qualify,'' so each write-in candidate shares the same identifier. We assume that if in a given contest, write-ins qualified and were consequently tabulated as individual candidates, the CVR published by the state would disambiguate them.} and are extracted from these files using a JSON parser written in Lean.

\begin{statute}{AS 15.15.350(g)(4)}
``ranking'' or ``ranked'' means the number assigned by a voter to a candidate to express the voter's choice for that candidate; a ranking of ``1'' is the highest ranking, followed by ``2,'' and then ``3,'' and so on;
\end{statute}

The form of the ballots used in Alaska allows voters to rank multiple candidates at the same ranking, no candidates at a ranking, or one candidate at multiple rankings. The definition of ``ranking'' as written cannot capture the last case if we interpret it as a function from candidates to rankings. Instead we must represent ballots as a list of lists: a ballot is a list of rankings, and each ranking contains a list of candidates. We provide a function \path{candsAt} for extracting the candidates ranked at a certain level on a given ballot. Moreover, the \textit{operative} parts of the statute are stated in terms of the candidates assigned to each ranking. For example, an overvote is a ranking assigned to more than one candidate (g)(3); a skipped ranking is a ranking whose candidate list is empty (g)(6); the tabulator must look at ``the next ranking'' if one is blank (e)(2). The \verb|ballotIndex| field is a unique identifier for each ballot and plays no role in tabulation. A given contest (whose type is named \path{Election}) is modeled as a list of \path{Ballot}s together with the list of candidates that may appear on a ballot. 

We assume that decisions about which marks count, how ambiguous marks are resolved, and which ballots are counted were made by the election administration before tabulation and are therefore outside the scope of our formalization. The correctness of the output of our tabulation function thus depends on the correctness of the CVR and the parser.

In Title 15, a round is defined as ``an instance of the sequence of voting tabulation in a general election.'' This serves as a pointer to (d), where each ``instance'' is parametrized by the set of continuing candidates and active ballots. We will discuss the choice to include the set of active ballots as a parameter in Section \ref{sec:inactive}.

\subsubsection{Overvotes}\label{sec:overvotes}  We take the definition of overvote in (g)(3) to mean that whether a ballot contains an overvote is a syntactic property of that ballot, determinable before tabulation. 
\begin{statute}{AS 15.15.350(g)(3)}
``overvote'' means an instance where a voter has assigned the same ranking to more than one candidate;
\end{statute}
The following Lean snippet contains the formal definition we wrote corresponding to (g)(3): 
\begin{MintedVerbatim}[commandchars=\\\{\}]
\PYG{k+kn}{def}\PYG{+w}{ }\PYG{n}{Ballot}\PYG{n+nb+bp}{.}\PYG{n}{IsOvervoteAt}\PYG{+w}{ }\PYG{o}{(}\PYG{n}{b}\PYG{+w}{ }\PYG{o}{:}\PYG{+w}{ }\PYG{n}{Ballot}\PYG{o}{)}\PYG{+w}{ }\PYG{o}{(}\PYG{n}{k}\PYG{+w}{ }\PYG{o}{:}\PYG{+w}{ }\PYG{n}{Nat}\PYG{o}{)}\PYG{+w}{ }\PYG{o}{:}\PYG{+w}{ }\PYG{k+kt}{Prop}\PYG{+w}{ }\PYG{o}{:=}
\PYG{+w}{  }\PYG{n+nb+bp}{∃}\PYG{+w}{ }\PYG{n}{c}\PYG{+w}{ }\PYG{n+nb+bp}{∈}\PYG{+w}{ }\PYG{n}{b}\PYG{n+nb+bp}{.}\PYG{n}{candsAt}\PYG{+w}{ }\PYG{n}{k}\PYG{o}{,}\PYG{+w}{ }\PYG{n+nb+bp}{∃}\PYG{+w}{ }\PYG{n}{c\PYGZsq{}}\PYG{+w}{ }\PYG{n+nb+bp}{∈}\PYG{+w}{ }\PYG{n}{b}\PYG{n+nb+bp}{.}\PYG{n}{candsAt}\PYG{+w}{ }\PYG{n}{k}\PYG{o}{,}\PYG{+w}{ }\PYG{n}{c}\PYG{+w}{ }\PYG{n+nb+bp}{≠}\PYG{+w}{ }\PYG{n}{c\PYGZsq{}}
\end{MintedVerbatim}
 
The text itself separates the definition of an overvote from its use. (g)(3) defines an overvote without mentioning continuing candidates, while (g)(2), immediately before it, brings in the continuing candidates only when it uses the term: an inactive ballot ``contains an overvote at the highest continuing ranking.'' Subsection (e)(1) uses similar wording. It's clear from reading either part of the statute that whether an overvote can make a ballot inactive depends on the round. 

\begin{statute}{AS 15.15.350(e)(1)}
a ballot containing an overvote shall be considered an inactive ballot once the overvote is encountered at the highest ranking for a continuing candidate;\footnote{Subsection (g)(2) words this condition differently: an inactive ballot ``contains an overvote at the highest continuing ranking.'' Rankings are not themselves continuing or defeated, so we read ``the highest continuing ranking'' as the highest ranking for a continuing candidate, and \path{IsHighestContinuingRank} matches both clauses.}
\end{statute}
\begin{MintedVerbatim}[commandchars=\\\{\}]
\PYG{k+kn}{def}\PYG{+w}{ }\PYG{n}{IsHighestContinuingRank}\PYG{+w}{ }\PYG{o}{(}\PYG{n}{cont}\PYG{+w}{ }\PYG{o}{:}\PYG{+w}{ }\PYG{n}{List}\PYG{+w}{ }\PYG{n}{Cand}\PYG{o}{)}\PYG{+w}{ }\PYG{o}{(}\PYG{n}{b}\PYG{+w}{ }\PYG{o}{:}\PYG{+w}{ }\PYG{n}{Ballot}\PYG{o}{)}\PYG{+w}{ }\PYG{o}{(}\PYG{n}{k}\PYG{+w}{ }\PYG{o}{:}\PYG{+w}{ }\PYG{n}{Nat}\PYG{o}{)}\PYG{+w}{ }\PYG{o}{:}\PYG{+w}{ }\PYG{k+kt}{Prop}\PYG{+w}{ }\PYG{o}{:=}
\PYG{+w}{  }\PYG{o}{(}\PYG{n+nb+bp}{∃}\PYG{+w}{ }\PYG{n}{c}\PYG{+w}{ }\PYG{n+nb+bp}{∈}\PYG{+w}{ }\PYG{n}{b}\PYG{n+nb+bp}{.}\PYG{n}{candsAt}\PYG{+w}{ }\PYG{n}{k}\PYG{o}{,}\PYG{+w}{ }\PYG{n}{c}\PYG{+w}{ }\PYG{n+nb+bp}{∈}\PYG{+w}{ }\PYG{n}{cont}\PYG{o}{)}\PYG{+w}{ }\PYG{n+nb+bp}{∧}\PYG{+w}{ }\PYG{n+nb+bp}{∀}\PYG{+w}{ }\PYG{n}{j}\PYG{+w}{ }\PYG{n+nb+bp}{\PYGZlt{}}\PYG{+w}{ }\PYG{n}{k}\PYG{o}{,}\PYG{+w}{ }\PYG{n+nb+bp}{∀}\PYG{+w}{ }\PYG{n}{c}\PYG{+w}{ }\PYG{n+nb+bp}{∈}\PYG{+w}{ }\PYG{n}{b}\PYG{n+nb+bp}{.}\PYG{n}{candsAt}\PYG{+w}{ }\PYG{n}{j}\PYG{o}{,}\PYG{+w}{ }\PYG{n}{c}\PYG{+w}{ }\PYG{n+nb+bp}{∉}\PYG{+w}{ }\PYG{n}{cont}

\PYG{k+kn}{def}\PYG{+w}{ }\PYG{n}{IsOvervoteAtHighestCont}\PYG{+w}{ }\PYG{o}{(}\PYG{n}{cont}\PYG{+w}{ }\PYG{o}{:}\PYG{+w}{ }\PYG{n}{List}\PYG{+w}{ }\PYG{n}{Cand}\PYG{o}{)}\PYG{+w}{ }\PYG{o}{(}\PYG{n}{b}\PYG{+w}{ }\PYG{o}{:}\PYG{+w}{ }\PYG{n}{Ballot}\PYG{o}{)}\PYG{+w}{ }\PYG{o}{:}\PYG{+w}{ }\PYG{k+kt}{Prop}\PYG{+w}{ }\PYG{o}{:=}
\PYG{+w}{  }\PYG{n+nb+bp}{∃}\PYG{+w}{ }\PYG{n}{k}\PYG{o}{,}\PYG{+w}{ }\PYG{n}{IsHighestContinuingRank}\PYG{+w}{ }\PYG{n}{cont}\PYG{+w}{ }\PYG{n}{b}\PYG{+w}{ }\PYG{n}{k}\PYG{+w}{ }\PYG{n+nb+bp}{∧}\PYG{+w}{ }\PYG{n}{b}\PYG{n+nb+bp}{.}\PYG{n}{IsOvervoteAt}\PYG{+w}{ }\PYG{n}{k}
\end{MintedVerbatim}

The phrase ``at the highest ranking for a continuing candidate'' is not defined independently in the statute or accompanying regulations, but formalizing (e)(1) and (g)(2) requires that we define it. The Lean definition \path{IsHighestContinuingRank} is meant to capture it. It holds of a rank $k$ on a ballot $b$ if there is a continuing candidate at that rank and none of the candidates ranked higher (i.e. at smaller numbers) are continuing. A faithful implementation of the procedure specified by the statute must check whether any candidate at an overvoted rank is \textit{continuing} before deciding whether to make the ballot inactive. This is the first point where the reproduction model differs from the statutory model. The two models use the same syntactic condition for overvotes while operationalizing those definitions differently. We call the statutory model’s overvote rule \emph{Stop}: an overvote makes a ballot permanently inactive once it is \textit{encountered at} the highest ranking for a continuing candidate. The reproduction model instead makes a ballot inactive when an overvote occurs \emph{at or above} that ranking, even if every candidate at the overvoted ranking has already been defeated. We call this rule \emph{Truncate} because it prevents the first overvoted ranking and all lower rankings from contributing a vote, which makes an overvoted ballot equivalent to a ballot consisting only of the marks above the overvote.\footnote{See \path{resultOf_toReproduction_eq_inactive_iff} for a formal statement and proof of this equivalence.} 

\subsubsection{Skipped rankings}\label{sec:skip} According to the statute,
\begin{statute}{AS 15.15.350(g)(6)}
``skipped ranking'' means a blank ranking on a ballot on which a voter has ranked another candidate at a subsequent ranking.
\end{statute} 
\noindent Like ``highest continuing ranking,'' the word ``blank'' does not get a formal definition anywhere in Alaska's statutes nor in accompanying election regulations. We will take ``blank at rank $k$'' to mean ``no mark was made at rank $k$.''\footnote{If in the original scan a mark was deemed ``ambiguous'' by the scanner and adjudicated to not count as indicative of intent to rank a candidate, we treat it as blank. This choice is supported by the structure of the CVRs, which do not record such marks in the \path{Modified} ballot records that record human-adjudicated ballot readings.} Section \ref{sec:empirical} discusses how the reproduction model treats skipped rankings, and Section \ref{sec:legal} discusses possible legal bases for the two competing interpretations.

\begin{MintedVerbatim}[commandchars=\\\{\}]
\PYG{k+kn}{def}\PYG{+w}{ }\PYG{n}{SkippedRankingAt}\PYG{+w}{ }\PYG{o}{(}\PYG{n}{b}\PYG{+w}{ }\PYG{o}{:}\PYG{+w}{ }\PYG{n}{Ballot}\PYG{o}{)}\PYG{+w}{ }\PYG{o}{(}\PYG{n}{k}\PYG{+w}{ }\PYG{o}{:}\PYG{+w}{ }\PYG{n}{Nat}\PYG{o}{)}\PYG{+w}{ }\PYG{o}{:}\PYG{+w}{ }\PYG{k+kt}{Prop}\PYG{+w}{ }\PYG{o}{:=}
\PYG{+w}{  }\PYG{n}{b}\PYG{n+nb+bp}{.}\PYG{n}{IsBlankAt}\PYG{+w}{ }\PYG{n}{k}\PYG{+w}{ }\PYG{n+nb+bp}{∧}\PYG{+w}{ }\PYG{n+nb+bp}{∃}\PYG{+w}{ }\PYG{n}{k\PYGZsq{}}\PYG{+w}{ }\PYG{n+nb+bp}{\PYGZlt{}}\PYG{+w}{ }\PYG{n}{b}\PYG{n+nb+bp}{.}\PYG{n}{ranks}\PYG{n+nb+bp}{.}\PYG{n}{length}\PYG{o}{,}\PYG{+w}{ }\PYG{n}{k}\PYG{+w}{ }\PYG{n+nb+bp}{\PYGZlt{}}\PYG{+w}{ }\PYG{n}{k\PYGZsq{}}\PYG{+w}{ }\PYG{n+nb+bp}{∧}\PYG{+w}{ }\PYG{n+nb+bp}{¬}\PYG{+w}{ }\PYG{n}{b}\PYG{n+nb+bp}{.}\PYG{n}{IsBlankAt}\PYG{+w}{ }\PYG{n}{k\PYGZsq{}}
\end{MintedVerbatim}

\noindent Like overvotes, skipped rankings are not defined relative to a round, while the inactivity condition to which they contribute is. 

\begin{statute}{AS 15.15.350(e)(2)}
if a ballot skips a ranking, then the election board shall count the next ranking. If the next ranking is another skipped ranking, the ballot shall be considered an inactive ballot once the second skipped ranking is encountered;
\end{statute}

The Lean theorem \path{readRankings_eq_iff} establishes that the procedural implementation (\path{readRankings}) corresponding to (e)(2) agrees with the declarative inactivity condition (\path{TwoSkips}) corresponding to (g)(2).

\subsubsection{Duplicate rankings}\label{sec:duplicates} Regulation 6 AAC 25.072 contains a rule that does not appear in the statute \cite{DuplicateRankings}. The form of the ballot makes it possible for a voter to mark the same candidate at more than one ranking, so any implementation of a tabulation procedure needs to decide how to treat such marks. 

\begin{statute}{6 AAC 25.072}
    ...a duplicate ranking occurs when a voter marks more than one ranking for the same candidate. The election board shall count a duplicate ranking as the highest ranking marked for the candidate and disregard the remaining rankings for that candidate.
\end{statute}

The first sentence is a syntactic definition, like the definition of an overvote: the predicate \texttt{HasDuplicateRanking} holds of a ballot that marks some candidate at two rankings, regardless of whether that candidate is continuing. The regulation leaves open whether disregarding a repeated mark simply prevents it from supplying a vote or erases it for other ballot checks. We adopt the former reading: repeated marks remain present, but cannot supply a later vote.\footnote{It is actually possible to achieve this semantics without any separate predicates or checking for duplicate marks. By the time tabulation reaches a repeated mark for a candidate, \textit{that candidate must already have been defeated}, so the mark wouldn't count anyway.} 

We call our interpretation \emph{Retain}: repeated candidate marks remain present when determining whether a ranking is blank or overvoted. We call the alternative \emph{Remove}: repeated marks are removed before applying either check, while ranking positions are preserved. The reproduction model uses a third treatment, which we call \emph{Hybrid}: repeated marks are disregarded when determining whether a ranking is blank but retained when detecting overvotes. Table \ref{tab:rules} contains a summary of the names we use to refer to different interpretations of overvotes and duplicate rankings. Section~\ref{sec:legal} assesses the competing readings.

\begin{table}[t]
\centering
\footnotesize
\begin{tabular}{@{}lp{0.65\linewidth}@{}}
\toprule
\textbf{Rule} & \textbf{Description} \\
\midrule
Truncate & Ignore the overvote and all lower rankings. \\
Stop & Ballot becomes inactive only when the overvote is at the level of the highest continuing candidate. \\
\midrule
Retain & Keep lower duplicates for overvote and skip detection. \\
Remove & Remove lower duplicates for both tests. \\
Hybrid & Remove for skips; retain for overvotes. \\
\bottomrule
\end{tabular}
\caption{Taxonomy of treatments of overvotes and duplicate rankings.}
\label{tab:rules}
\end{table}

\subsubsection{Inactive ballots}\label{sec:inactive}

A ballot may become inactive for any of three reasons: it meets the round-parametrized overvote condition, it is exhausted, or it contains two sequential\footnote{We take ``sequential'' to mean ``consecutive.''} skipped rankings before the highest continuing ranking.

\begin{statute}{AS 15.15.350(g)(2)}
``inactive ballot'' means a ballot that is no longer tabulated, either in whole or in part, by the division because it does not rank any continuing candidate,\footnote{Alaska uses ``exhausted'' to mean ``ranks no continuing candidates.'' Some jurisdictions use ``exhausted'' for ``inactive.''} contains an overvote at the highest continuing ranking, or contains two or more sequential skipped rankings before its highest continuing ranking;

\end{statute}

We represent the inactivity condition described in (g)(2) by \path{MeetsInactivityCondition}, which is the disjunction of the three conditions \path{Exhausted}, \path{TwoSkips}, and \path{IsOvervoteAtHighestCont}. 

The first subtlety in deciding inactivity is the parametrization of inactivity-due-to-overvotes by the set of continuing candidates as discussed in Section \ref{sec:overvotes}. The second has to do with the \textit{persistence} of inactivity. A ballot which was inactive in one round under \path{Exhausted} or \path{TwoSkips} will remain inactive under one of those conditions in every possible future round. The theorem \path{twoSkips_or_exhausted_of_subset} is the formal statement of this fact.

\begin{figure}[H]
  \centering
  \small
  \setlength{\tabcolsep}{3pt}
  \renewcommand{\arraystretch}{1.0}
  \begin{tabular}{@{}l*{5}{c}@{}}
    \ballothead
    \ballotline{\textbf{Begich, Nick}}{1}{0}{0}{0}{0}
    \ballotline{\textbf{Bye, Chris}}{0}{1}{0}{0}{0}
    \ballotline{\textbf{Palin, Sarah}}{0}{0}{1}{0}{0}
    \ballotline{\textbf{Peltola, Mary S.}}{0}{0}{0}{1}{0}
    \ballotline{Write-in:}{0}{1}{0}{0}{0}
    \bottomrule
  \end{tabular}

  \caption{A ballot reconstructed from Alaska's 2022 general-election cast vote record (ballot 241786)
    \cite{alaskadivisionofelectionsCastVoteRecord2022}.}
  \label{fig:overvote-ballot}
\end{figure}

The predicate \path{IsOvervoteAtHighestCont} does not, once it becomes true of a ballot, stay true of that ballot. Consider the ballot represented in Figure \ref{fig:overvote-ballot}. Write-ins do not qualify so are not continuing. If Begich was defeated first, then in round two, this ballot would satisfy \path{IsOvervoteAtHighestCont}, since Bye is the highest-ranked continuing candidate and the ballot contains a syntactic overvote at that rank. Now imagine that Bye is defeated in round 2. Then the highest continuing candidate is Palin, and there is no overvote at her rank, so \path{MeetsInactivityCondition} is false in this round when it was true in an earlier round.

A naive implementation of the tabulation algorithm would simply recompute at each round which ballots are inactive. This implementation would be ``stateless'' with respect to the inactive ballots because the tabulator does not need to keep track of extra state beyond which candidates are continuing.

This would be incorrect. The statute says that inactive ballots are ``\textit{no longer} tabulated,'' which we take to mean that once a ballot becomes inactive, it should not become active in any later round and thereby contribute to vote totals in later rounds. The Division likewise explains that ballots are excluded from tabulation once they become inactive \cite[``Votes to include in threshold calculation'']{alaskadivisionofelectionsElectionResults2022}. Instead, each round carries the ballots that are still active, and defeating a candidate removes the ballots that become inactive.\footnote{The reproduction model never has to make this choice. It reads nothing below a ballot's first overvote, so a ballot that reaches an overvote never counts again.}

\begin{MintedVerbatim}[commandchars=\\\{\}]
\PYG{k+kn}{def}\PYG{+w}{ }\PYG{n}{next}\PYG{+w}{ }\PYG{o}{(}\PYG{n}{r}\PYG{+w}{ }\PYG{o}{:}\PYG{+w}{ }\PYG{n}{Round}\PYG{o}{)}\PYG{+w}{ }\PYG{o}{(}\PYG{n}{d}\PYG{+w}{ }\PYG{o}{:}\PYG{+w}{ }\PYG{n}{Cand}\PYG{o}{)}\PYG{+w}{ }\PYG{o}{:}\PYG{+w}{ }\PYG{n}{Round}\PYG{+w}{ }\PYG{k+kn}{where}
\PYG{+w}{  }\PYG{n}{continuing}\PYG{+w}{ }\PYG{o}{:=}\PYG{+w}{ }\PYG{n}{r}\PYG{n+nb+bp}{.}\PYG{n}{continuing}\PYG{n+nb+bp}{.}\PYG{n}{erase}\PYG{+w}{ }\PYG{n}{d}
\PYG{+w}{  }\PYG{n}{ballots}\PYG{+w}{ }\PYG{o}{:=}\PYG{+w}{ }\PYG{n}{r}\PYG{n+nb+bp}{.}\PYG{n}{ballots}\PYG{n+nb+bp}{.}\PYG{n}{filter}\PYG{+w}{ }\PYG{k}{fun}\PYG{+w}{ }\PYG{n}{b}\PYG{+w}{ }\PYG{n+nb+bp}{=\PYGZgt{}}
\PYG{+w}{    }\PYG{n}{decide}\PYG{+w}{ }\PYG{o}{(}\PYG{n+nb+bp}{¬}\PYG{+w}{ }\PYG{n}{MeetsInactivityCondition}\PYG{+w}{ }\PYG{o}{(}\PYG{n}{r}\PYG{n+nb+bp}{.}\PYG{n}{continuing}\PYG{n+nb+bp}{.}\PYG{n}{erase}\PYG{+w}{ }\PYG{n}{d}\PYG{o}{)}\PYG{+w}{ }\PYG{n}{b}\PYG{o}{)}
\end{MintedVerbatim}

Finally we can say how votes are assigned to candidates within each round. 

\begin{statute}{AS 15.15.350(d)}
When counting ballots in a general election, the election board shall initially tabulate each validly cast ballot as one vote for the highest-ranked continuing candidate on that ballot or as an inactive ballot...
\end{statute}

We define the ``result'' of a ballot with respect to the set of continuing candidates as either inactive or a vote for a candidate $C$.  The predicate \path{IsBallotResult} describes a ballot's result for a fixed list of continuing candidates; it does not record inactivity in earlier rounds. Actual tabulation also depends on whether the ballot remains among the round's active ballots. The history-dependent predicate \path{IsBallotResultAfter} captures this distinction: a ballot removed in an earlier round remains inactive even if its remaining rankings would otherwise support a continuing candidate.

From here we define a few helpful functions, like \texttt{votesFor c} which computes the number of ballots whose \path{BallotResult} is a \texttt{voteFor c} relative to the set of continuing candidates and \path{votes} which does the same thing relative to a \path{Round} object.

\subsection{Allowable Outcomes}
We represent the outcomes permitted under the statute as a relational specification, that is, as a predicate stating whether a given output is allowed. Permitted election outcomes are stated in terms of permitted \textit{round} outcomes.

\begin{statute}{AS 15.15.350(d)(1)}
if two or fewer continuing candidates remain, the candidate with the greatest number of votes is elected and the tabulation is complete; otherwise, the tabulation continues under (2) of this subsection;
\end{statute}
\begin{MintedVerbatim}[commandchars=\\\{\}]
\PYG{k+kn}{def}\PYG{+w}{ }\PYG{n}{Round}\PYG{n+nb+bp}{.}\PYG{n}{Elects}\PYG{+w}{ }\PYG{o}{(}\PYG{n}{r}\PYG{+w}{ }\PYG{o}{:}\PYG{+w}{ }\PYG{n}{Round}\PYG{o}{)}\PYG{+w}{ }\PYG{o}{(}\PYG{n}{c}\PYG{+w}{ }\PYG{o}{:}\PYG{+w}{ }\PYG{n}{Cand}\PYG{o}{)}\PYG{+w}{ }\PYG{o}{:}\PYG{+w}{ }\PYG{k+kt}{Prop}\PYG{+w}{ }\PYG{o}{:=}
\PYG{+w}{  }\PYG{n}{r}\PYG{n+nb+bp}{.}\PYG{n}{continuing}\PYG{n+nb+bp}{.}\PYG{n}{length}\PYG{+w}{ }\PYG{n+nb+bp}{≤}\PYG{+w}{ }\PYG{l+m+mi}{2}\PYG{+w}{ }\PYG{n+nb+bp}{∧}\PYG{+w}{ }\PYG{n}{c}\PYG{+w}{ }\PYG{n+nb+bp}{∈}\PYG{+w}{ }\PYG{n}{r}\PYG{n+nb+bp}{.}\PYG{n}{continuing}\PYG{+w}{ }\PYG{n+nb+bp}{∧}
\PYG{+w}{    }\PYG{n+nb+bp}{∀}\PYG{+w}{ }\PYG{n}{c\PYGZsq{}}\PYG{+w}{ }\PYG{n+nb+bp}{∈}\PYG{+w}{ }\PYG{n}{r}\PYG{n+nb+bp}{.}\PYG{n}{continuing}\PYG{o}{,}\PYG{+w}{ }\PYG{n}{r}\PYG{n+nb+bp}{.}\PYG{n}{votes}\PYG{+w}{ }\PYG{n}{c\PYGZsq{}}\PYG{+w}{ }\PYG{n+nb+bp}{≤}\PYG{+w}{ }\PYG{n}{r}\PYG{n+nb+bp}{.}\PYG{n}{votes}\PYG{+w}{ }\PYG{n}{c}
\end{MintedVerbatim}

A round $R$ may elect a candidate $C$ if and only if there are two or fewer continuing candidates,\footnote{If at least two candidates are continuing after write-ins that do not qualify are removed, there will never be fewer than two candidates in any round. See \path{two_le_length_roundAfter_continuing} for a proof.} of which $C$ is one, and if $C$ receives at least as many votes as the other candidate, if there is any. A tie between the final two candidates is resolved under the procedures referenced in (e)(3), which we do not model; instead \path{Elects} is true of both candidates, since either outcome is permissible. 

\begin{statute}{AS 15.15.350(d)(2)}
if the candidate with the fewest votes is defeated, votes cast for the defeated candidate shall cease counting for the defeated candidate...
\end{statute}
\begin{minipage}{\linewidth}
\begin{MintedVerbatim}[commandchars=\\\{\}]
\PYG{k+kn}{def}\PYG{+w}{ }\PYG{n}{Round}\PYG{n+nb+bp}{.}\PYG{n}{CanDefeat}\PYG{+w}{ }\PYG{o}{(}\PYG{n}{r}\PYG{+w}{ }\PYG{o}{:}\PYG{+w}{ }\PYG{n}{Round}\PYG{o}{)}\PYG{+w}{ }\PYG{o}{(}\PYG{n}{d}\PYG{+w}{ }\PYG{o}{:}\PYG{+w}{ }\PYG{n}{Cand}\PYG{o}{)}\PYG{+w}{ }\PYG{o}{:}\PYG{+w}{ }\PYG{k+kt}{Prop}\PYG{+w}{ }\PYG{o}{:=}
\PYG{+w}{  }\PYG{l+m+mi}{3}\PYG{+w}{ }\PYG{n+nb+bp}{≤}\PYG{+w}{ }\PYG{n}{r}\PYG{n+nb+bp}{.}\PYG{n}{continuing}\PYG{n+nb+bp}{.}\PYG{n}{length}\PYG{+w}{ }\PYG{n+nb+bp}{∧}\PYG{+w}{ }\PYG{n}{d}\PYG{+w}{ }\PYG{n+nb+bp}{∈}\PYG{+w}{ }\PYG{n}{r}\PYG{n+nb+bp}{.}\PYG{n}{continuing}\PYG{+w}{ }\PYG{n+nb+bp}{∧}
\PYG{+w}{    }\PYG{n+nb+bp}{∀}\PYG{+w}{ }\PYG{n}{d\PYGZsq{}}\PYG{+w}{ }\PYG{n+nb+bp}{∈}\PYG{+w}{ }\PYG{n}{r}\PYG{n+nb+bp}{.}\PYG{n}{continuing}\PYG{o}{,}\PYG{+w}{ }\PYG{n}{r}\PYG{n+nb+bp}{.}\PYG{n}{votes}\PYG{+w}{ }\PYG{n}{d}\PYG{+w}{ }\PYG{n+nb+bp}{≤}\PYG{+w}{ }\PYG{n}{r}\PYG{n+nb+bp}{.}\PYG{n}{votes}\PYG{+w}{ }\PYG{n}{d\PYGZsq{}}
\end{MintedVerbatim}

\end{minipage}

Read literally, (d)(2) does not say that the candidate with the fewest votes \emph{is} defeated; it only says what happens to that candidate's votes if they are. The rule itself is left implicit, but the rest of the section leaves little doubt about what it should be: (e)(3) resolves a tie ``between two candidates with the fewest votes'' by lot ``to determine which candidate is defeated,'' which only makes sense if the candidate with the fewest votes is the one defeated.\footnote{Subsection (e)(3) addresses only a tie between \emph{two} last-place candidates; the statute says nothing about a tie among three or more. \path{CanDefeat} handles multi-way ties, which we take to be the evident intent.} We read (d)(2) accordingly. 

Thus, a round $R$ may defeat a candidate $C$ if and only if at least three candidates are continuing, $C$ is one of them, and no continuing candidate has fewer votes than $C$. The first condition comes from (d)(1), under which a round with two or fewer continuing candidates ends the tabulation instead. The predicate \path{CanDefeat} is extended recursively to a sequence of rounds with the predicate \path{ValidDefeats}, which asserts that starting from a given \path{Round}, a given list of candidates is an acceptable defeat order. 

Note that this interpretation builds in a treatment of ties for last place. Since ties are resolved by lot, it is always permissible, depending on the outcome of the coin toss, to eliminate any of the candidates tied. Thus it is permissible to defeat any candidate with the lowest number of votes in a round. 

Finally, we tie the permitted outcomes of individual rounds into the permitted outcomes of the entire tabulation.

\begin{MintedVerbatim}[commandchars=\\\{\}]
\PYG{k+kn}{def}\PYG{+w}{ }\PYG{n}{Elected}\PYG{+w}{ }\PYG{o}{(}\PYG{n}{e}\PYG{+w}{ }\PYG{o}{:}\PYG{+w}{ }\PYG{n}{Election}\PYG{o}{)}\PYG{+w}{ }\PYG{o}{(}\PYG{n}{defeated}\PYG{+w}{ }\PYG{o}{:}\PYG{+w}{ }\PYG{n}{List}\PYG{+w}{ }\PYG{n}{Cand}\PYG{o}{)}\PYG{+w}{ }\PYG{o}{(}\PYG{n}{c}\PYG{+w}{ }\PYG{o}{:}\PYG{+w}{ }\PYG{n}{Cand}\PYG{o}{)}\PYG{+w}{ }\PYG{o}{:}\PYG{+w}{ }\PYG{k+kt}{Prop}\PYG{+w}{ }\PYG{o}{:=}
\PYG{+w}{  }\PYG{o}{(}\PYG{n}{IsMajorityWinner}\PYG{+w}{ }\PYG{o}{(}\PYG{n}{initialRound}\PYG{+w}{ }\PYG{n}{e}\PYG{o}{)}\PYG{n+nb+bp}{.}\PYG{n}{continuing}\PYG{+w}{ }\PYG{o}{(}\PYG{n}{initialRound}\PYG{+w}{ }\PYG{n}{e}\PYG{o}{)}\PYG{n+nb+bp}{.}\PYG{n}{ballots}\PYG{+w}{ }\PYG{n}{c}\PYG{+w}{ }\PYG{n+nb+bp}{∧}\PYG{+w}{ }\PYG{n}{defeated}\PYG{+w}{ }\PYG{n+nb+bp}{=}\PYG{+w}{ }\PYG{o}{[])}\PYG{+w}{ }\PYG{n+nb+bp}{∨}
\PYG{+w}{  }\PYG{o}{(}\PYG{n}{NoMajority}\PYG{+w}{ }\PYG{n}{e}\PYG{+w}{ }\PYG{n+nb+bp}{∧}\PYG{+w}{ }\PYG{o}{(}\PYG{n}{initialRound}\PYG{+w}{ }\PYG{n}{e}\PYG{o}{)}\PYG{n+nb+bp}{.}\PYG{n}{ValidDefeats}\PYG{+w}{ }\PYG{n}{defeated}\PYG{+w}{ }\PYG{n+nb+bp}{∧}\PYG{+w}{ }\PYG{o}{(}\PYG{n}{roundAfter}\PYG{+w}{ }\PYG{n}{e}\PYG{+w}{ }\PYG{n}{defeated}\PYG{o}{)}\PYG{n+nb+bp}{.}\PYG{n}{Elects}\PYG{+w}{ }\PYG{n}{c}\PYG{o}{)}
\end{MintedVerbatim}

This predicate is true of an election, sequence of defeated candidates, and winning candidate $C$ if and only if either (i) $C$ was the majority winner in the first-round check and the list of defeated candidates is empty, or (ii) there was no majority winner, the defeat sequence is valid according to \path{ValidDefeats}, and the round that follows after defeating all the candidates in the defeat sequence \path{Elects} $C$. 

\subsection{Implementation}\label{sec:impl}
\begin{figure}[H]
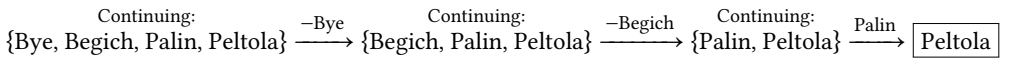

\centering
\resizebox{0.95\linewidth}{!}{%
$\overset{\text{Continuing:}}{\left\{\text{Bye, Begich, Palin, Peltola}\right\}}
\xrightarrow{-\text{Bye}}
\overset{\text{Continuing:}}{\left\{\text{Begich, Palin, Peltola}\right\}}
\xrightarrow{-\text{Begich}}
\overset{\text{Continuing:}}{\left\{\text{Palin, Peltola}\right\}}
\xrightarrow{\text{Palin}}
\boxed{\text{Peltola}}$
}
\caption{The \texttt{RCVTree} representing the 2022 general election for U.S. Representative.}
\label{fig:rcvtree}
\end{figure}

The specification states which outcomes are permitted but does not compute them. Since we do not model the procedures that resolve ties, an election can have more than one permitted outcome, and the function \path{tabulate} returns all of them as an \path{RCVTree}: a finite rooted tree whose edges and leaves are labeled by candidates. Figure~\ref{fig:rcvtree} represents the defeat order of a real Alaska contest as an \path{RCVTree}. The full Lean definitions of \path{RCVTree} and \path{tabulate} can be found in Appendix~\ref{app:listings}.

A node has more than one child only when two or more candidates are tied for the fewest votes. The statute resolves such ties by lot under (e)(3), so the outcome of tiebreaks cannot be determined from ballots. To represent this, we do what in computational social choice is called \emph{parallel-universes tiebreaking} (PUT) \cite{conitzerPreferenceFunctionsThat, wangPracticalAlgorithmsMultiStage2019} and report all PUT winners: a candidate is a PUT winner if some way of breaking the ties makes that candidate win. Appendix~\ref{app:tiebreak} gives an example of an \path{RCVTree} with a tie.

Reading the edge labels along a path from the root to a leaf gives the order in which the candidates are defeated. The last label on each \textit{path} (not the label of the leaf node) is the candidate who is not elected in the final round. Under (d)(1), that candidate is not  \emph{defeated}; the other candidate is \emph{elected}. When a candidate has a majority in the initial tabulation, the tree is a single leaf and the path is empty.

The function \path{tabulate} produces exactly the winners permitted by our interpretation of the statute. That fact is formalized as the following theorem: 
  
\begin{theorem}[\path{tabulate_correct}]\label{thm:tabulate_correct}
  For every election $e$, list of candidates $D$ and candidate $c$, the statutory model permits $c$ to be elected after the candidates in $D$ are defeated in order iff the \path{RCVTree} computed by \path{tabulate e} has a path to a leaf labeled by $c$ whose labels, other than the last, are $D$.
  \end{theorem}

Theorem \ref{thm:tabulate_correct} is a result about the relationship between the function and the specification gleaned from the statute. We can specialize it to state theorems that convey ``this CVR under this interpretation elects this candidate,'' but that requires importing entire CVRs into the \textit{theorem} as a definition rather than using it as input to an executable function.\footnote{After downloading the CVRs, run \texttt{lake exe alaska <export-dir> [contest-id]} to tabulate a selected contest. For example, \texttt{lake exe alaska cvr/2022-special 69} computes the tabulation for the 2022 special U.S.\ House election.} To make election-specific proofs tractable, we group identical rankings into ballot profiles: the 192,289 ballots cast in the 2022 special election contain just 691 distinct ranking patterns. We prove that tabulating the \textit{profiles} gives the same result as tabulating the full set of represented ballots. For all 115 contests, we establish the statutory and reproduction models' complete tabulation trees with computations checked by Lean's kernel. 

Assuming the CVRs are published and parsed correctly, both models produce the same single-path (i.e. no ties) \texttt{RCVTree} in every contest. Together with Theorem~\ref{thm:tabulate_correct}, the statutory-model equalities establish that these represent the unique permitted winner and defeat order for each contest. We also prove that the reproduction model's per-round ballot disposition counts are equal to those from the published reports. Thus, both the election outcomes and their agreement with published round-by-round results are expressed as formally checked propositions.

\section{Empirical Results}\label{sec:empirical}
After writing our \emph{statutory model} (Section \ref{sec:statutory}), which we developed independently of the published results and the vendor, we constructed a variation of our code that reproduces the published round-by-round results of all 22 multi-round contests. The Division's report for the 2022 special general election \cite[p.~1]{alaskadivisionofelectionsRCVTabulation2022} lists some of its tabulator's settings, and we took those as given. We varied the remaining rules individually, including how to treat a duplicate ranking, and whether a ranking that marks only a write-in counts as skipped. We compared each variant with every number in the 22 round-by-round reports: each candidate's votes and the number of blank, exhausted and overvoted ballots in every round. Among the rule combinations we tested, exactly one produced a result that matched the published tallies (see Appendix~\ref{app:identify}).

We refer to this model as the \emph{reproduction model}. The reproduction model reuses much of the formal representations developed for the statutory model to enable easy comparisons between the two readings. We use the models to identify individual ballots that they count differently, for example a ballot that one model counts toward a candidate in a given round while the other treats it as inactive.

We analyzed 115 contests using the published cast vote records from Alaska's August 2022 special general election, November 2022 general election, and November 2024 general election
\cite{alaskadivisionofelectionsCastVoteRecord2022a,
alaskadivisionofelectionsCastVoteRecord2022,
alaskadivisionofelectionsCastVoteRecord2024}. We compare our tabulations with the corresponding official results
\cite{alaskadivisionofelectionsElectionResults2022a,
alaskadivisionofelectionsElectionResults2022,
alaskadivisionofelectionsElectionResults2024}.

The Division describes the CVR as the tabulator's input but notes that it excludes ballots counted only by hand \cite[``Cast Vote Record'']{alaskadivisionofelectionsElectionResults2022}. Our comparison assumes that the published CVRs contain the ballots used for the corresponding ranked-choice reports. No write-in candidates qualified in any of the 115 analyzed contests. We also assume that the same set of ballot-reading rules was applied to all of the reported contests.

The reproduction model combines Truncate for overvotes with Hybrid for duplicate rankings (refer to Table \ref{tab:rules} for a summary of the rules). Under Hybrid, repeated candidate marks are disregarded when determining whether a ranking is blank but retained when detecting overvotes. The Division's 2022 general-election results glossary says that write-in rankings are not treated as blank or skipped merely because they are for write-ins \cite[``Exclude unresolved write-ins'']{alaskadivisionofelectionsElectionResults2022}. However, the reproduction model treats write-in marks as subject to Hybrid like other repeated marks. 

Across 115 contests, 75 ballot records count for a candidate under the statutory model in at least one round in which they are inactive under the reproduction model. That is, the reproduction model counts certain ballots in fewer rounds than does the statutory model. These differences occur in nine contests with published multi-round reports and in the 2022 governor contest, where one ballot counts for Walker/Drygas under the statutory model but is inactive under the reproduction model in the initial tabulation. The greatest numbers of differences were found in the 2022 special-election U.S. House contest (29), the 2024 U.S. House contest (17), and the 2022 general-election U.S. House contest (14). The comparisons below establish how the models treat individual records, but their treatment by the deployed software is inferred from the reproduction model.

Overall, 34,836 ballot records contain duplicate rankings and 9,535 contain overvotes. To separate the two mechanisms for treating these ballot errors, we compare both models with an intermediate model combining Truncate with Retain. With Retain fixed, replacing Stop with Truncate makes three ballots inactive in the final round that would otherwise count for a candidate: two for Murkowski in the 2022 general-election U.S. Senate contest and one for Palin in the U.S. House contest. With Truncate fixed, replacing Retain with Hybrid makes 72 additional ballots inactive in at least one round in which they would otherwise count for a candidate. These two groups are disjoint and together account for all 75 differences between the statutory and reproduction models.\footnote{Each ballot–contest record is counted once, even if its treatment differs in several rounds.} Both models elect the same winners and follow the same defeat orders in all 115 contests. Their final margins differ by at most seven votes, attained in the 2022 special election for U.S. Representative. As such, simply comparing winners would not reveal the differences in the models. 

The fact that the two models agree on winners does not make these differences immaterial. Alaska's published final-round tallies include margins as small as seven votes (2022 House District 15). Although this contest does not have ballot-treatment differences between our models, the rules require justification independently of the observed margins of victory. 

Going forward, we will focus on the 2022 general-election contest for U.S. Representative as an illustrative example because it exhibits both discrepancy mechanisms: 13 ballots differ because of duplicate treatment and one because of overvote treatment. The 2022 special election has more differing ballots, but all 29 differences arise from duplicate treatment.

\begin{figure}[t]
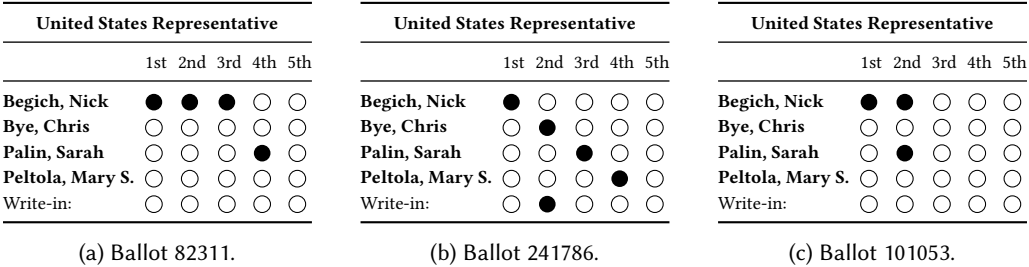

  \centering
  \scriptsize
  \setlength{\tabcolsep}{2pt}
  \renewcommand{\arraystretch}{1.2}
  \captionsetup[subfigure]{font=scriptsize}

  \begin{subfigure}[t]{0.32\linewidth}
    \centering
    \begin{tabular}{@{}l*{5}{c}@{}}
      \ballothead
      \ballotline{\textbf{Begich, Nick}}{1}{1}{1}{0}{0}
      \ballotline{\textbf{Bye, Chris}}{0}{0}{0}{0}{0}
      \ballotline{\textbf{Palin, Sarah}}{0}{0}{0}{1}{0}
      \ballotline{\textbf{Peltola, Mary S.}}{0}{0}{0}{0}{0}
      \ballotline{Write-in:}{0}{0}{0}{0}{0}
      \bottomrule
    \end{tabular}

    \caption{Ballot 82311.}
    \label{bal:a}
  \end{subfigure}\hfill
  \begin{subfigure}[t]{0.32\linewidth}
    \centering
    \begin{tabular}{@{}l*{5}{c}@{}}
      \ballothead
      \ballotline{\textbf{Begich, Nick}}{1}{0}{0}{0}{0}
      \ballotline{\textbf{Bye, Chris}}{0}{1}{0}{0}{0}
      \ballotline{\textbf{Palin, Sarah}}{0}{0}{1}{0}{0}
      \ballotline{\textbf{Peltola, Mary S.}}{0}{0}{0}{1}{0}
      \ballotline{Write-in:}{0}{1}{0}{0}{0}
      \bottomrule
    \end{tabular}

    \caption{Ballot 241786.}
    \label{bal:b}
  \end{subfigure}\hfill
  \begin{subfigure}[t]{0.32\linewidth}
    \centering
    \begin{tabular}{@{}l*{5}{c}@{}}
      \ballothead
      \ballotline{\textbf{Begich, Nick}}{1}{1}{0}{0}{0}
      \ballotline{\textbf{Bye, Chris}}{0}{0}{0}{0}{0}
      \ballotline{\textbf{Palin, Sarah}}{0}{1}{0}{0}{0}
      \ballotline{\textbf{Peltola, Mary S.}}{0}{0}{0}{0}{0}
      \ballotline{Write-in:}{0}{0}{0}{0}{0}
      \bottomrule
    \end{tabular}

    \caption{Ballot 101053.}
    \label{bal:c}
  \end{subfigure}

  \caption{Ballot records reconstructed from the published CVR for Alaska's November 2022 general election
    \cite{alaskadivisionofelectionsCastVoteRecord2022}.}
  \label{fig:house2022-ballots}
\end{figure}

Figure~\ref{fig:house2022-ballots} and Table~\ref{tab:house2022-ballot-rounds} illustrate the two mechanisms by which the ballots can be interpreted differently. Ballot~(\subref{bal:a}) illustrates the duplicate-ranking mechanism affecting 13 ballots. Ballot~(\subref{bal:b}) is the sole differing ballot in this contest attributable to overvote treatment. Ballot~(\subref{bal:c}), discussed in Section~\ref{sec:dupl-ballot}, illustrates why removing repeated marks produces a different result from either retaining them or applying Hybrid.

The statutory model counts all 14 ballots for some candidate in every round. The reproduction model assigns them to the same candidates until they become inactive: 13 in the final round and one in round 2. The latter ballot (in class (a)) ranks Bye first through fourth and Peltola fifth. In the final round, the statutory model assigns 10 of the 14 ballots to Palin and four to Peltola. Peltola’s margin of victory is consequently reduced by six votes: she receives 24,786 in the final round under the statutory model, compared with 24,792 under the reproduction model.

\subsection{Reproduction vs. Statutory Model}
Both models defeat Bye and then Begich, leaving Palin and Peltola in the final round. Table~\ref{tab:house2022-ballot-rounds} traces ballots (a) and (b) through that sequence.

\begin{table}[t]
  \small
  \begin{tabular}{@{}llccc@{}}
    \toprule
    ballot & model & round 1 & round 2 & round 3 \\
    & & & (Bye defeated) & (Begich defeated) \\
    \midrule
    (a) 82311 & statutory & Begich & Begich & Palin \\
    & reproduction & Begich & Begich & inactive (two skipped rankings) \\
    \midrule
    (b) 241786 & statutory & Begich & Begich & Palin \\
    & reproduction & Begich & Begich & inactive (overvote at rank 2) \\
    \bottomrule
  \end{tabular}
  \caption{The results of ballots (\subref{bal:a}) and (\subref{bal:b}) in each round, under the statutory and reproduction models.}
  \label{tab:house2022-ballot-rounds}
\end{table}

Table \ref{tab:house2022-ballot-rounds} isolates the two mechanisms of difference. In Ballot~(\subref{bal:a}), Hybrid treats the repeated marks for Begich as blank, while Retain preserves them and transfers the vote to Palin after Begich is defeated. In Ballot~(\subref{bal:b}), Truncate ignores all preferences at and below the overvote, while Stop passes over that rank because neither Bye nor the write-in candidate is continuing when Begich is defeated. This allows the vote to reach Palin.  

This result depends on the defeat order. Had Begich been defeated before Bye, ranking 2 would have become the ballot’s highest continuing ranking while Bye remained continuing. Stop would then have made the ballot permanently inactive, including after Bye’s subsequent defeat (\path{overvoteBallot_defeat_order}).\footnote{ Stop's dependence on the order of candidate defeats raises a separate question about whether or not that is a desirable property of a treatment of overvotes. We do not attempt to answer that question here.}

It is clear that the voter strictly prefers Bye to Palin and Palin to Peltola, but under Truncate these kinds of ascertainable preferences are never included in the final tabulation.

\subsection{Hybrid vs. Retain and  Remove}\label{sec:dupl-ballot}
Ballot~(\subref{bal:a}) distinguishes Retain from Hybrid and Remove; Ballot~(\subref{bal:c}) distinguishes Remove from Hybrid and Retain. Alvarez-Rivera previously identified the duplicate treatment needed to reproduce the 2022 results in \cite{alvarez-riveraRankedChoiceVoting2023}.

Ballot~(\subref{bal:c}) shows why removing repeated marks before both checks also changes the result. Its second ranking contains a repeated mark for Begich alongside a mark for Palin. Hybrid retains the second mark for Begich when detecting overvotes, so the reproduction model makes the ballot inactive when Begich is defeated. Remove would erase that repeated mark, leaving Palin alone at ranking 2 and allowing the ballot to count for her. Retain agrees with Hybrid on this ballot.

Holding Truncate fixed, replacing Hybrid with Retain allows 13 ballots in this contest to count for a candidate in a round where the reproduction model makes them inactive. Replacing Hybrid with Remove does so for 15 ballots. Across all 115 contests, the corresponding counts are 72 and 46 ballot records. Neither alternative reproduces all the published reports: with Retain, at least one candidate tally or inactivity-category count differs from the published report in 11 of the 22 multi-round contests; with Remove, this occurs in eight. Table~\ref{tab:multiround} in Appendix~\ref{app:contests} presents these results for each contest.

\section{Legal Analysis}\label{sec:legal}
Alaska courts employ a ``sliding-scale'' approach to statutory interpretation: the clearer the statutory language, the stronger the evidence of contrary legislative purpose or intent must be. Statutory provisions are interpreted in context, with attention to their language, structure, and purpose \cite{AlaskaDemocraticParty2025}.

Alaska courts have also consistently favored reasonable interpretations of election statutes that preserve votes rather than unnecessarily disenfranchise voters. In \textit{Carr v. Thomas}, the Alaska Supreme Court emphasized that courts should favor reasonable statutory constructions that avoid disenfranchisement \cite[pp.~626--627]{CarrThomas1978}. In \textit{Miller v. Treadwell}, the court reaffirmed this principle, holding that election statutes should be interpreted to preserve voters' ascertainable choices rather than exclude votes for technical errors \cite[pp.~869--870]{MillerTreadwell2010}.

\subsection{Overvotes}
The statutory text favors Stop: both (e)(1) and (g)(2) condition inactivity on an overvote at the highest continuing ranking. Truncate blocks lower preferences even when no candidate at the overvoted rank remains continuing. 

The strongest argument in favor of Truncate rests on the word ``encountered'' in subsection (e)(1). A tabulator reading rankings in order encounters an overvote before reaching any lower preference. However, the word is immediately qualified by ``at the highest ranking for a continuing candidate.'' Reading ``encountered'' as sufficient by itself would deprive that qualification of operative significance.\footnote{Treating ``encounter'' as sufficient resembles our interpretation of ``disregard'' as \textit{not} erasing marks (Section~\ref{sec:duplicates}). However, our reading is that the regulation to disregard duplicate marks is redundant with the statute: a candidate must already be defeated by the time their second ranking could be reached. By contrast, ``highest continuing ranking'' has additional semantic content that would be lost without handling it separately.} The same reading must also account for subsection (g)(2), which restates the highest-continuing-ranking condition without the word ``encountered.'' The Division's non-mandatory explanatory materials do not resolve this interpretive question. Its results glossary \cite[entries ``Skip overvoted rankings'' and ``Exhausted'']{alaskadivisionofelectionsElectionResults2022} states that an overvote invalidates all lower rankings, while its definition of an exhausted ballot retains the statutory qualification concerning the highest continuing ranking. The glossary thus announces Truncate without explaining how it follows from the qualified inactivity condition presented simultaneously.

A complication arises from another provision enacted by Ballot Measure 2. AS 15.58.020(a)(13) requires the Division to inform voters in the official election pamphlet that their \textit{ballot} (not just certain rankings on that ballot) ``will not be counted if you assign the same ranking to more than one candidate for the same office.'' Taken literally, this statement describes a rule more restrictive than either Stop or Truncate: both models count a ballot's unambiguous higher preferences even when an overvote occurs at a lower ranking. The pamphlet language appears to be inconsistent with \textit{both} operative tabulation provisions rather than clearly favoring either interpretation. Although it forms part of Ballot Measure 2 and bears on its interpretation, it  does not independently determine the conditions under which an overvote renders a ballot inactive.

The Alaska Supreme Court considered an analogous discrepancy from the same ballot measure in \textit{Alaska Democratic Party v. Beecher}. That case concerned whether the Division could successively replace multiple candidates who withdrew from the general-election ballot. The court observed that AS 15.58.020(a)(13), as enacted by Ballot Measure 2, prescribed pamphlet language suggesting that only one candidate could be replaced, whereas the operative replacement provision admitted a broader interpretation. The court declined to treat the language from the pamphlet as controlling, observing that the drafters had not used the same language in the provision actually governing candidate replacement. It concluded that the contradiction, at most, further highlighted the statute's ambiguity \cite[slip op. p.~18]{AlaskaDemocraticParty2025}.

Moreover, if Truncate is what the authors had in mind when writing the statute, it would have been straightforward for them to specify this. By 2018, Colorado's election rules expressly provided that an overvote invalidates the overvoted ranking and all lower rankings \cite[rule~26.8.1]{ColoradoElectionRules2018}. Colorado had specified and implemented Truncate before Alaska adopted Ballot Measure 2. RCTab, an open-source tabulator, also documents an ``Exhaust immediately'' overvote option in its version 2.0 user guide \cite[``Voter Error Rules Tab'']{QuickStartRCTab}. Because such language already existed and was in use for similar purposes but was not used in the statute, we should be cautious about assuming this was the intended interpretation.

The distinction between Stop and Truncate also has consequences for ballot inclusion. Our formal results establish that, for any set of continuing candidates, every ballot that Truncate counts for a candidate is counted for the same candidate under Stop, and that some ballots are counted under Stop but not under Truncate.\footnote{See \path{isVoteFor_of_isVoteFor_truncateAtOvervote} for the inclusion and \path{isVoteFor_truncateAtOvervote_ssubset_isVoteFor} for a ballot showing that it is strict.} Thus, Truncate excludes some votes that Stop would count, while Stop excludes none that Truncate would count. Although \textit{Miller} concerned the validity of write-in votes rather than overvotes, its preference for preserving ascertainable voter choices provides an additional reason to favor Stop, provided that the statutory text admits both interpretations.

\subsection{Duplicate Rankings}

The reproduction model raises a separate interpretive question concerning duplicate rankings. Under 6 AAC 25.072, when a voter ranks the same candidate more than once, the election board must count the candidate at their highest ranking and ``disregard the remaining rankings for that candidate.'' The regulation does not specify how this instruction interacts with the statutory definitions of overvotes and skipped rankings.

The reproduction model follows neither Remove nor Retain. Instead, Hybrid follows Remove when detecting skipped rankings and follows Retain when detecting overvotes. Consequently, a lower duplicate ranking can be absent for one statutory test while remaining present for another.

Different treatment in different contexts is not inherently impermissible. Alaska courts have nevertheless emphasized that statutory provisions should be interpreted together to produce a ``harmonious whole.'' In \textit{Peninsula Marketing Association v. State}, the Alaska Supreme Court rejected an interpretation that applied equivalent statutory language differently across related provisions without a sufficient basis for that distinction \cite[p.~921]{PeninsulaMarketingState1991}.

One possible justification would distinguish between disregarding a ranking for purposes of assigning a vote to a candidate and retaining the physical mark for purposes of detecting an overvote. But that distinction does not explain why the same physical mark is treated as ``blank'' (see Section \ref{sec:skip}) when detecting skipped rankings. Neither the regulation's instruction to disregard remaining rankings for that candidate nor the statutory definitions expressly establish this distinction.

Alternative treatments are documented in other ranked-choice voting systems. With exhaustion on repeated candidates disabled, RCTab's version 2.0 documentation describes repeated rankings as excluded-candidate rankings rather than skipped rankings. Its documented sequence of checks retains the candidates at a rank for overvote detection, consistent with our interpretation~\cite{QuickStartRCTab,RCTabSoftwareDesignV2}. By contrast, Multnomah County, Oregon, expressly provides that a candidate's highest ranking that is not an overvote is retained, while lower rankings are treated as skipped~\cite{multnomahcountyCode2025}. This resembles Remove, but actually specifies a completely separate interpretation of duplicate rankings. Dominion's version 5.17-CO::4 user guide\footnote{We could not locate publicly available Alaska-specific Dominion technical documentation. We therefore consult the Colorado-specific guides, which document the same vendor's Democracy Suite RCV tabulation software, including its configurable treatment of malformed ballots. These guides concern a different version and configuration, so we use them to assess what Dominion publicly documents.} lists configurable treatments of overvotes and skipped rankings \cite[pp.~71--73]{dominionResultsTallyReporting2023}. Its automated test deck generator explicitly supports ballots with duplicate rankings \cite[pp.~32--33]{dominionAutomatedTestDeck2023}, but neither guide specifies how duplicates are treated during tabulation or the order in which these checks are performed. It is therefore unclear whether Hybrid reflects a deliberate interpretive choice by the Division or an undocumented consequence of the tabulator's implementation.

Moreover, Hybrid's treatment is not merely a matter of interpretive consistency. Our formal results establish that with Truncate fixed and for any set of continuing candidates, every ballot that Hybrid counts for a candidate is counted for the same candidate under Remove and under Retain, and that each of Remove and Retain counts some ballots that Hybrid does not.\footnote{See \path{isVoteFor_truncateAtOvervote_dropDuplicates_of_isVoteFor_toReproduction} and \path{isVoteFor_truncateAtOvervote_of_isVoteFor_toReproduction} for the inclusions, and \path{isVoteFor_toReproduction_ssubset_isVoteFor_truncateAtOvervote_dropDuplicates} and \path{isVoteFor_toReproduction_ssubset_isVoteFor_truncateAtOvervote} for ballots showing that they are strict.} Hybrid therefore excludes votes that either alternative would count, without counting any votes that either alternative would exclude.

As with Truncate, this additional exclusion weighs against Hybrid under Alaska's preference for preserving ascertainable voter choices. However, this does not necessarily settle which of Remove or Retain is preferable: each represents a possible interpretation of ``disregard,'' and the preference for counting votes does not override statutory requirements.

\section{Related Work}\label{sec:related}
Several previous studies have analyzed Alaska's RCV elections using cast vote records. Alvarez-Rivera \cite{alvarez-riveraRankedChoiceVoting2023} identified the treatment of duplicate rankings necessary to reproduce the published results of Alaska's 2022 elections. Pacuit \cite{pacuit2023electionanalysis} investigated the 2022 special general election, including the treatment of overvotes and skipped rankings, although the resulting tabulations do not reproduce the published totals. Graham-Squire and McCune \cite{graham-squireMathematicalAnalysis20222022} reproduced the election totals and analyzed the election's social choice properties, including its failure to elect the Condorcet winner and its susceptibility to voting paradoxes. Our work extends these investigations by constructing a model that reproduces the published round-by-round tallies and comparing it with an independently formalized interpretation of Alaska's statutes.

The formal verification of ranked-choice voting algorithms is not new. Previous work has developed a framework for voting theory in Lean 3 and used it to verify properties of the Split Cycle voting method \cite{hollidayVotingTheoryLean2021}, and has verified IRV procedures in Dafny \cite{buttnerVerificationEVotingAlgorithms2025}, Rocq \cite{verityElectronicVoteCounting2016, ghaleModularFormalisationVerification2018, pattinsonVoteCountingMathematical2015}, and other systems \cite{dawsonMachineCheckedReasoningComplex2015,goel2018irv}. However, much of this work begins with a specified voting rule and asks whether an implementation satisfies it. In our case, deciding what the rule should be is itself part of the problem.

A particularly close precedent is Free \& Fair's \path{formal-rcv} \cite{FormalrcvFormalizedDevelopments}, which predates Alaska's adoption of RCV. It develops a formalization of RCV in Rocq based on San Francisco's election law. It explicitly identifies ambiguity in the statutory treatment of overvotes and adopts an interpretation informed by observed tabulation practices. Our work follows this approach to statutory formalization, but distinguishes the statutory specification from a model of observed tabulation behavior, allowing us to identify and analyze discrepancies between them in actual election results. The top-level implementation correctness proof in \path{formal-rcv} remains incomplete.

Catala \cite{merigouxCatalaProgrammingLanguage2021} provides a language for translating legal requirements into executable code while preserving their relationship to the legal text. We share this goal, but chose Lean to support both our particular formalization and the broader FVDemocracy project for several reasons. Catala does not support explicit recursion, which we use to directly express the statute's repeated rounds of tabulation. Moreover, in future work we plan to develop abstract frameworks for ballot interpretation (of course including alternative treatments of overvotes and duplicate rankings) in the same language as the statutory models, and prove relationships between those abstractions and particular statutory interpretations. Lean allows us to combine executable tabulation, statutory specifications, and general mathematical reasoning within one development.

In Slovenia, Bauer, Ber\v{c}i\v{c}, and Zagorc produced an independent validation of the 2022 parliamentary elections \cite{bauer2023volilni}. Through independent computations, their work identified errors in election software that initially resulted in the wrong candidates being reported as elected. They also investigated ambiguities and deficiencies in the governing electoral law.

Finally, Delemazure and Peters \cite{delemazureGeneralizingInstantRunoff2024} study extensions of IRV that permit voters to express indifference by ranking multiple candidates equally. Their results provide a voting-theoretic perspective on the preferences expressed by overvoted rankings, which existing U.S. RCV tabulation rules generally do not treat as valid expressions of indifference.

\section{Conclusion and Future Work}

This case study connects a formal interpretation of Alaska's election rules, a verified implementation thereof, and a comparison with the published election results. Making these interpretations explicit can help legislators, administrators, and developers resolve disagreements before they affect elections. 

We propose two main directions for future work. First, similar issues arise in other jurisdictions. Maine's RCV statute also refers to an overvote at the highest continuing ranking, while its regulations invalidate rankings at or below an overvote \cite{maineRCVRule, maineRCVStatute}. The work done for Alaska's elections can be repeated to better understand RCV rules around the country.

Second, we plan to develop a general framework for ballot interpretation rules. This will characterize relationships between rules and interpretations, generate distinguishing examples, and identify when these differences can affect election outcomes. Such a framework also needs a formal account of ascertainable voter preferences.

\begin{acks}
We used large language models to assist with proof construction and manuscript preparation, as described in Appendix~\ref{app:llms}.

We would like to thank Wes Holliday for suggesting the investigation that led to this work and for his guidance and feedback throughout the project.
\end{acks}

\bibliography{references}
\newpage
\appendix

\section{Lean Listings}\label{app:listings}

\begin{listing}[t]
\centering
\begin{minipage}[t]{0.32\linewidth}\vspace{0pt}
\begin{MintedVerbatim}[commandchars=\\\{\}]
\PYG{k+kn}{inductive}\PYG{+w}{ }\PYG{n}{Cand}\PYG{+w}{ }\PYG{k+kn}{where}
\PYG{+w}{  }\PYG{n+nb+bp}{|}\PYG{+w}{ }\PYG{n}{writeIn}\PYG{+w}{ }\PYG{o}{(}\PYG{n}{c}\PYG{+w}{ }\PYG{o}{:}\PYG{+w}{ }\PYG{n}{Nat}\PYG{o}{)}
\PYG{+w}{  }\PYG{n+nb+bp}{|}\PYG{+w}{ }\PYG{n}{regular}\PYG{+w}{ }\PYG{o}{(}\PYG{n}{c}\PYG{+w}{ }\PYG{o}{:}\PYG{+w}{ }\PYG{n}{Nat}\PYG{o}{)}
\end{MintedVerbatim}

\end{minipage}\hspace{0.08\linewidth}
\begin{minipage}[t]{0.32\linewidth}\vspace{0pt}
\begin{MintedVerbatim}[commandchars=\\\{\}]
\PYG{k+kn}{structure}\PYG{+w}{ }\PYG{n}{Ballot}\PYG{+w}{ }\PYG{k+kn}{where}
\PYG{+w}{  }\PYG{n}{ballotIndex}\PYG{+w}{ }\PYG{o}{:}\PYG{+w}{ }\PYG{n}{Nat}
\PYG{+w}{  }\PYG{n}{ranks}\PYG{+w}{ }\PYG{o}{:}\PYG{+w}{ }\PYG{n}{List}\PYG{+w}{ }\PYG{o}{(}\PYG{n}{List}\PYG{+w}{ }\PYG{n}{Cand}\PYG{o}{)}
\end{MintedVerbatim}

\end{minipage}

\medskip
\begin{minipage}[t]{0.32\linewidth}\vspace{0pt}
\begin{MintedVerbatim}[commandchars=\\\{\}]
\PYG{k+kn}{structure}\PYG{+w}{ }\PYG{n}{Election}\PYG{+w}{ }\PYG{k+kn}{where}
\PYG{+w}{  }\PYG{n}{ballots}\PYG{+w}{ }\PYG{o}{:}\PYG{+w}{ }\PYG{n}{List}\PYG{+w}{ }\PYG{n}{Ballot}
\PYG{+w}{  }\PYG{n}{cands}\PYG{+w}{ }\PYG{o}{:}\PYG{+w}{ }\PYG{n}{List}\PYG{+w}{ }\PYG{n}{Cand}
\PYG{+w}{  }\PYG{n}{cands\PYGZus{}unique}\PYG{+w}{ }\PYG{o}{:}\PYG{+w}{ }\PYG{n}{cands}\PYG{n+nb+bp}{.}\PYG{n}{Nodup}
\end{MintedVerbatim}

\end{minipage}\hspace{0.08\linewidth}
\begin{minipage}[t]{0.32\linewidth}\vspace{0pt}
\begin{MintedVerbatim}[commandchars=\\\{\}]
\PYG{k+kn}{structure}\PYG{+w}{ }\PYG{n}{Round}\PYG{+w}{ }\PYG{k+kn}{where}
\PYG{+w}{  }\PYG{n}{continuing}\PYG{+w}{ }\PYG{o}{:}\PYG{+w}{ }\PYG{n}{List}\PYG{+w}{ }\PYG{n}{Cand}
\PYG{+w}{  }\PYG{n}{ballots}\PYG{+w}{ }\PYG{o}{:}\PYG{+w}{ }\PYG{n}{List}\PYG{+w}{ }\PYG{n}{Ballot}

\end{MintedVerbatim}

\end{minipage}

\caption{Candidates, ballots, elections, and rounds.}
\label{lst:datamodel}
\end{listing}

Section~\ref{sec:impl} describes the tree returned by \path{tabulate}. This appendix gives the definitions of the tree and of \path{tabulate}.

\begin{MintedVerbatim}[commandchars=\\\{\}]
\PYG{k+kn}{inductive}\PYG{+w}{ }\PYG{n}{RCVTree}\PYG{+w}{ }\PYG{k+kn}{where}
\PYG{+w}{  }\PYG{n+nb+bp}{|}\PYG{+w}{ }\PYG{n}{winner}\PYG{+w}{ }\PYG{o}{:}\PYG{+w}{ }\PYG{n}{Cand}\PYG{+w}{ }\PYG{n+nb+bp}{→}\PYG{+w}{ }\PYG{n}{RCVTree}
\PYG{+w}{  }\PYG{n+nb+bp}{|}\PYG{+w}{ }\PYG{n}{branch}\PYG{+w}{ }\PYG{o}{:}\PYG{+w}{ }\PYG{n}{List}\PYG{+w}{ }\PYG{o}{(}\PYG{n}{Cand}\PYG{+w}{ }\PYG{n+nb+bp}{×}\PYG{+w}{ }\PYG{n}{RCVTree}\PYG{o}{)}\PYG{+w}{ }\PYG{n+nb+bp}{→}\PYG{+w}{ }\PYG{n}{RCVTree}
\end{MintedVerbatim}

\noindent A leaf \texttt{winner c} records that candidate $c$ is elected. A node \texttt{branch children} has, for each pair $(d, t)$ in \path{children}, an edge labeled $d$ leading to the subtree $t$. In a round with three or more continuing candidates, $d$ is a candidate the round \path{CanDefeat}s, and $t$ contains every permitted continuation of the tabulation after $d$ is defeated. In the final round, $d$ is the candidate who is not elected, and $t$ is the leaf for the elected candidate. The \textit{type} \path{RCVTree} doesn't enforce this correspondence to an election. We compute the outcome of an election as an \path{RCVTree} using the following function:

\begin{MintedVerbatim}[commandchars=\\\{\}]
\PYG{k+kn}{def}\PYG{+w}{ }\PYG{n}{tabulate}\PYG{+w}{ }\PYG{o}{(}\PYG{n}{e}\PYG{+w}{ }\PYG{o}{:}\PYG{+w}{ }\PYG{n}{Election}\PYG{o}{)}\PYG{+w}{ }\PYG{o}{:}\PYG{+w}{ }\PYG{n}{RCVTree}\PYG{+w}{ }\PYG{o}{:=}
\PYG{+w}{  }\PYG{k}{let}\PYG{+w}{ }\PYG{n}{cont₁}\PYG{+w}{ }\PYG{o}{:=}\PYG{+w}{ }\PYG{n}{initialCands}\PYG{+w}{ }\PYG{n}{e}
\PYG{+w}{  }\PYG{k}{match}\PYG{+w}{ }\PYG{n}{majorityWinner}\PYG{+w}{ }\PYG{n}{cont₁}\PYG{+w}{ }\PYG{n}{e}\PYG{n+nb+bp}{.}\PYG{n}{ballots}\PYG{+w}{ }\PYG{k}{with}
\PYG{+w}{  }\PYG{n+nb+bp}{|}\PYG{+w}{ }\PYG{n}{none}\PYG{+w}{ }\PYG{n+nb+bp}{=\PYGZgt{}}
\PYG{+w}{    }\PYG{n}{tabulateRounds}\PYG{+w}{ }\PYG{n}{cont₁}\PYG{+w}{ }\PYG{o}{(}\PYG{n}{e}\PYG{n+nb+bp}{.}\PYG{n}{ballots}\PYG{n+nb+bp}{.}\PYG{n}{filter}\PYG{+w}{ }\PYG{k}{fun}\PYG{+w}{ }\PYG{n}{b}\PYG{+w}{ }\PYG{n+nb+bp}{=\PYGZgt{}}
\PYG{+w}{        }\PYG{n}{decide}\PYG{+w}{ }\PYG{o}{(}\PYG{n+nb+bp}{¬}\PYG{+w}{ }\PYG{n}{MeetsInactivityCondition}\PYG{+w}{ }\PYG{n}{cont₁}\PYG{+w}{ }\PYG{n}{b}\PYG{o}{))}
\PYG{+w}{  }\PYG{n+nb+bp}{|}\PYG{+w}{ }\PYG{n}{some}\PYG{+w}{ }\PYG{n}{c}\PYG{+w}{ }\PYG{n+nb+bp}{=\PYGZgt{}}\PYG{+w}{ }\PYG{n}{RCVTree}\PYG{n+nb+bp}{.}\PYG{n}{winner}\PYG{+w}{ }\PYG{n}{c}
\end{MintedVerbatim}

It computes an \path{RCVTree} by following the rules laid out in (d). First, the function \path{initialCands} checks if the write-ins qualify, and removes them from the list of continuing candidates if they do not. Then it checks if any candidate wins a majority of first-choice votes in the first round. If so, the resulting \path{RCVTree} is just the leaf node labeled by that candidate. If not, it follows \path{tabulateRounds}, which executes the recursive tabulation procedure described in the statute, stopping recursion when two candidates remain and electing the one that has more votes in that round.

We also prove that the first-round majority check and the final round do not need to be treated separately. Simply running the intermediate rounds from the initial set of continuing candidates until there is one remaining is guaranteed to elect the same candidate(s) as \texttt{tabulate}. We prove this as \path{exists_elected_iff_mem_winners_recurseRounds}.

\section{An Election Decided by Tiebreaking}\label{app:tiebreak}

\begin{figure}[H]
  \centering
  \begin{minipage}[c]{0.4\linewidth}
  \centering
  \begin{tabular}{cl}
  \toprule
  Ballots & Ranking \\
  \midrule
  3 & $A$ \\
  2 & $B$, $C$ \\
  2 & $C$, $A$ \\
  \bottomrule
  \end{tabular}
  \end{minipage}%
  \begin{minipage}[c]{0.4\linewidth}
  \centering
  \begin{tikzpicture}[
      level distance=11mm,
      level 1/.style={sibling distance=20mm},
      inner/.style={circle, fill, minimum size=4pt, inner sep=0pt},
      leaf/.style={draw, inner sep=3pt}]
    \node[inner] {}
      child { node[inner] {}
        child { node[leaf] {$C$} edge from parent node[left] {$A$} }
        edge from parent node[above left] {$B$} } 
      child { node[inner] {}
        child { node[leaf] {$A$} edge from parent node[right] {$B$} }
        edge from parent node[above right] {$C$} };
  \end{tikzpicture}
  \end{minipage}
  \caption{An election whose outcome is determined by tiebreaking.}
  \label{fig:app-rcvtree}
  \end{figure}
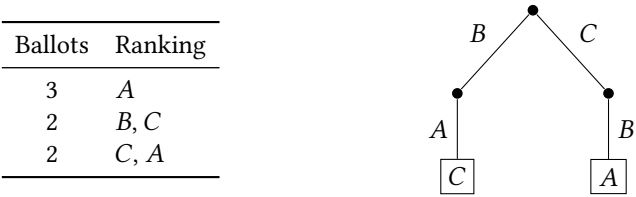

Figure~\ref{fig:app-rcvtree} shows an example of a contest with three candidates and seven ballots. In the initial tabulation, $A$ has three votes and $B$ and $C$ have two each, so no candidate has a majority. $B$ and $C$ tie for the fewest votes, so either may be defeated. If $B$ is defeated, $B$'s ballots count for $C$, who wins the final round with four votes to $A$'s three. If $C$ is defeated, $C$'s ballots count for $A$, who wins the final round with five votes to $B$'s two. The resolution of the tie therefore determines the winner, so this election's associated tree records both possibilities. In the example, the paths are $[B, A]$ to $C$ and $[C, B]$ to $A$, so the specification permits $C$ to be elected after $B$ is defeated, and $A$ to be elected after $C$ is defeated.

\section{Identifying the Reproduction Model}\label{app:identify}

The Division's report for the 2022 special general election lists its tabulator's settings \cite[p.~1]{alaskadivisionofelectionsRCVTabulation2022}: unresolved write-ins are excluded, overvoted rankings are not skipped, and a ballot is exhausted when two or more rankings are skipped. We took these settings as a starting point for the reproduction model. We changed one of the remaining rules at a time and compared each variant with the 22 published round-by-round reports. Table~\ref{tab:variants} gives the number of reports each variant matches. Only the reproduction model matches all 22.

\begin{table}[h]
  \small
\begin{tabular}{@{}lcc@{}}
\toprule
change from the reproduction model & \shortstack{\# of reports matched\\on candidate votes}
& \shortstack{\# of reports matched\\on all counts}  \\
\midrule
none (reproduction model: Truncate, Hybrid) & 22 & 22 \\
Stop instead of Truncate & 20 & 19 \\
Retain instead of Hybrid & 13 & 11 \\
Remove instead of Hybrid & 14 & 14 \\
a ranking marking only write-ins is blank & 12 & 12 \\
a ranking with no continuing candidate is blank & 0 & 0 \\
one skipped ranking makes a ballot inactive & 0 & 0 \\
skipped rankings never make a ballot inactive & 0 & 0 \\
statutory model (Stop, Retain) & 13 & 11 \\
\bottomrule
\end{tabular}
  \caption{Number of published round-by-round reports matched by the reproduction model and by variants that each change one of its rules.}
  \label{tab:variants}
\end{table}

 A variant matches a report ``on candidate votes'' if every candidate's tally agrees in every round, and ``on all counts'' if the numbers of blank, exhausted and overvoted ballots also agree. These do not differ very often, but they just represent ballots being counted as inactive for different reasons by different rule sets. 

\section{Multi-Round Contests}\label{app:contests}

Table~\ref{tab:multiround} lists the 22 contests that went past the initial round, and the 2022 governor contest. For each contest it gives the final-round margin under the statutory and reproduction models, the number of ballot records the two models count differently in some round, divided by the rule responsible, and whether each of four rule combinations matches the published report. Among the 93 contests decided in the initial round, the models count a ballot differently only in the 2022 governor contest, where one ballot differs because of duplicate treatment.

\begin{table}[h]
  \scriptsize
  \setlength{\tabcolsep}{3pt}
\begin{tabular}{@{}lrcrrrrcccc@{}}
\toprule
& & & \multicolumn{2}{c}{final margin} & \multicolumn{2}{c}{\shortstack{differing ballots \\ between T/H and S/Ret}} & \multicolumn{4}{c}{matches report} \\
\cmidrule(lr){4-5}\cmidrule(lr){6-7}\cmidrule(l){8-11}
contest & ballots & rounds & stat. & repr. & overvote & duplicate & S/H & T/Ret & T/Rem & S/Ret \\
\midrule
\multicolumn{11}{@{}l}{\emph{2022 special}} \\
\quad U.S. House & 192,289 & 2 & 5,233 & 5,240 & 0 & 29 & -- & -- & -- & -- \\
\multicolumn{11}{@{}l}{\emph{2022 general}} \\
\quad U.S. Senate & 266,797 & 3 & 18,799 & 18,796 & 2 & 6 & -- & -- & -- & -- \\
\quad U.S. House & 266,797 & 3 & 24,786 & 24,792 & 1 & 13 & -- & -- & -- & -- \\
\quad Senate D & 16,236 & 2 & 1,010 & 1,010 & 0 & 0 & $\cmark$ & $\cmark$ & $\cmark$ & $\cmark$ \\
\quad Senate E & 17,437 & 2 & 1,932 & 1,932 & 0 & 0 & $\cmark$ & $\cmark$ & -- & $\cmark$ \\
\quad Senate N & 13,776 & 2 & 1,813 & 1,813 & 0 & 0 & $\cmark$ & -- & $\cmark$ & -- \\
\quad House 11 & 8,190 & 2 & 113 & 112 & 0 & 1 & $\cmark$ & -- & $\cmark$ & -- \\
\quad House 15 & 7,709 & 2 & 7 & 7 & 0 & 0 & $\cmark$ & $\cmark$ & $\cmark$ & $\cmark$ \\
\quad House 18 & 2,297 & 2 & 78 & 77 & 0 & 1 & $\cmark$ & -- & $\cmark$ & -- \\
\quad House 28 & 7,098 & 3 & 1,184 & 1,184 & 0 & 0 & $\cmark$ & $\cmark$ & $\cmark$ & $\cmark$ \\
\quad House 30 & 8,135 & 2 & 778 & 778 & 0 & 0 & $\cmark$ & $\cmark$ & $\cmark$ & $\cmark$ \\
\quad House 31 & 5,330 & 2 & 515 & 515 & 0 & 0 & $\cmark$ & $\cmark$ & $\cmark$ & $\cmark$ \\
\quad House 34 & 7,693 & 2 & 813 & 813 & 0 & 0 & $\cmark$ & $\cmark$ & $\cmark$ & $\cmark$ \\
\multicolumn{11}{@{}l}{\emph{2024 general}} \\
\quad U.S. House & 335,853 & 3 & 7,878 & 7,876 & 0 & 17 & $\cmark$ & -- & -- & -- \\
\quad Senate D & 20,138 & 2 & 1,688 & 1,688 & 0 & 0 & $\cmark$ & $\cmark$ & -- & $\cmark$ \\
\quad Senate F & 17,967 & 2 & 934 & 934 & 0 & 0 & $\cmark$ & $\cmark$ & $\cmark$ & $\cmark$ \\
\quad Senate L & 20,340 & 2 & 1,952 & 1,953 & 0 & 1 & $\cmark$ & -- & -- & -- \\
\quad House 6 & 11,798 & 2 & 489 & 488 & 0 & 1 & $\cmark$ & -- & $\cmark$ & -- \\
\quad House 28 & 8,915 & 2 & 12 & 12 & 0 & 0 & $\cmark$ & $\cmark$ & $\cmark$ & $\cmark$ \\
\quad House 36 & 9,382 & 3 & 917 & 915 & 0 & 2 & $\cmark$ & -- & -- & -- \\
\quad House 38 & 3,707 & 3 & 137 & 137 & 0 & 0 & $\cmark$ & -- & $\cmark$ & -- \\
\quad House 40 & 3,170 & 2 & 533 & 533 & 0 & 0 & $\cmark$ & $\cmark$ & $\cmark$ & $\cmark$ \\
\multicolumn{11}{@{}l}{\emph{decided in the initial round}} \\
\quad 2022 Governor & 266,693 & 1 & 68,697 & 68,697 & 0 & 1 & & & & \\
\midrule
total & & & & & 3 & 72 &  &  &  &  \\
\bottomrule
\end{tabular}
  \caption{The 22 multi-round contests in Alaska from 2022-2024 plus the one contest decided in the initial round in which the models count a ballot differently.}
  \label{tab:multiround}
\end{table}
``Margin'' is the winner's lead over the runner-up in the final round. The differing ballots are those the statutory and reproduction models count differently in some round: ``overvote'' ballots differ between Stop and Truncate with Retain fixed, and ``duplicate'' ballots differ between Retain and Hybrid with Truncate fixed. A rule combination (S = Stop, T = Truncate, H = Hybrid, Ret = Retain, Rem = Remove) matches the report if every candidate's tally and the numbers of blank, exhausted and overvoted ballots agree in every round. The reproduction model (T/H) matches every report. S/Ret is the statutory model.

\section{Use of Large Language Models}\label{app:llms}
We used large language models (primarily Claude Opus 5.5 as a coding agent, occasionally GPT-6 Astra as a chat model) to help construct Lean proofs and some theorem statements, functions and properties after specifying them in natural language. This kind of human-computer pair programming is a particularly promising use of LLMs in formal verification: the proofs must be correct, but humans need not write them, since the Lean kernel can check them regardless of how they were produced. Moreover, as long as the human coder can judge the correctness of the declarations and step in to fix them when needed, LLMs can have a hand in writing them. What requires human judgment is deciding what to formalize and whether the definitions and statements actually capture the intended properties of the system. We either wrote or personally reviewed all major formal definitions and theorem statements. Automating the construction of proofs and intermediate declarations therefore lets us spend our time on the parts of formal verification that actually require human judgment, without sacrificing guarantees of correctness.

We also used LLMs to assist with revising portions of the manuscript, managing citations, and to identify potentially relevant legal precedent, including Alaska Supreme Court decisions concerning statutory interpretation and election law. We remain responsible for the interpretations and arguments presented in this paper.
\end{document}